\documentclass{article}
\usepackage{graphicx}

\begin{document}







\title{Four basic symmetry types in the universal 7-cluster
structure of 143 complete bacterial genomic sequences}

\author{
\large{Gorban Alexander} \\ {\it Centre for Mathematical
Modelling, University of Leicester, UK}
\\ \large{Popova Tatyana} \\ {\it Institute of Computational Modelling SB RAS, Krasnoyarsk,
Russia}
 \\ \large{Zinovyev Andrey} \footnote{To whom correspondence should be addressed: {\it
zinovyev@ihes.fr}} \\ {\it Institut des Hautes Études
Scientifiques, Bures-sur-Yvette, France}
 }

\maketitle

\abstract{ Coding information is the main source of heterogeneity
(non-randomness) in the sequences of bacterial genomes. This
information can be naturally modeled by analysing cluster structures
in the ``in-phase'' triplet distributions of relatively short
genomic fragments (200-400bp). We found a universal 7-cluster
structure in bacterial genomic sequences and explained its
properties. We show that codon usage of bacterial genomes is a
multi-linear function of their genomic G+C-content with high
accuracy. Based on the analysis of 143 completely sequenced
bacterial genomes available in Genbank in August 2004, we show that
there are four ``pure'' types of the 7-cluster structure observed.
All 143 cluster animated 3D-scatters are collected in a database and
is made available on our web-site:
http://www.ihes.fr/$\sim$zinovyev/7clusters. The finding can be
readily introduced into any software for gene prediction, sequence
alignment or bacterial genomes classification }

\section{Introduction}

The bacterial genomes are compact genomes: most of the sequence
contains coding information. Hence any statistical study of
bacterial genomic sequence will detect coding information as the
main source of heterogeneity (non-randomness). This is confirmed
by mining sequences ``from scratch'', without use of any
biological information, using entropic or Hidden Markov Modeling
(HMM) statistical approaches (for examples, see \cite{Claverie98},
\cite{Baldi00}, \cite{Bernaola}, \cite{Prum02}). All these methods
can be seen as specific clustering of relatively short genomic
fragments of length in the range 200-400bp comparable to the
average length of a coding information piece.

Surprisingly, not much is known about the properties of the
cluster structure itself, independently on the gene recognition
problems, where it is implicitly used since long time ago (see,
for example, early paper \cite{Borodovsky93} about famous GENMARK
gene-predictor, or \cite{Salzberg98} about GLIMMER approach). Only
recently the structure was described explicitly. In
\cite{Gorban03}, \cite{GorbanGeneRecPrep01}, \cite{Zinovyev02} and
\cite{GorbanOpSys03} the structure was visualized in the
64-dimensional space of non-overlapping triplet distributions for
several genomes. Also the same dataset was visualized in
\cite{GorbanVizPrep01} and \cite{GorbanIJCNN03} using non-linear
principal manifolds. In \cite{Zhang03} several particular cases of
this structure were observed in the context of the Z-curve
methodology in the 9-dimensional space of Z-coordinates: it was
claimed that the structure has interesting flower-like pattern but
can be observed only for GC-rich genomes. This is somehow in
contradiction with the results of \cite{Zinovyev02}, published
before, where the same flower-like picture was demonstrated for
AT-rich genome of {\it Helicobacter pylori}. This fact shows that
this simple and basic structure is far from being completely
understood and described.

The problem can be stated in the following way: there is a set of
genomic fragments of length 100-1000 bp representing a genome
almost uniformly. There are various ways to produce this set, for
example, by sliding window with a given step of sliding (in this
case sequence assembly is not generally needed), or it might be a
full set of ORFs (in this case one needs to know the assembled
sequence). We construct a distribution of points in a
multidimensional space of statistics calculated on the fragments
and study the cluster structure of this distribution. The
following questions arise: What is the number of clusters? What is
the character of their mutual locations? Is there a ``thin
structure'' in the clusters? How the structure depends on the
properties of genomic sequence, can we predict it?

Every fragment can be characterized by a ``frequency dictionary''
of short words (see examples in \cite{Gorban93}, \cite{Gorban98},
\cite{Gorban00}, \cite{Karlin98}). For our purposes we use
frequencies of non-overlapping triplets, counted from the first
basepair of a fragment. Thus every fragment is a point in
64-dimensional space of triplet frequencies. This choice is not
unique, moreover, we use dimension reduction techniques to
simplify this description and take the essential features. The
cluster structure we are going to describe is universal in the
sense that it is observed in any bacterial genome and with any
type of statistics which takes into account coding phaseshifts.
The structure is basic in the sense that it is revealed in the
analysis in the first place, serving as the principal source of
sequence non-randomness. In \cite{Gorban03},
\cite{GorbanGeneRecPrep01}, \cite{Zhang03}, \cite{Zinovyev02} it
was shown that even simple clustering methods like K-Means or
Fuzzy K-Means give good results in application of the structure to
gene-finding.

\begin{figure}
\centering{
\includegraphics[width=120mm, height=100mm]{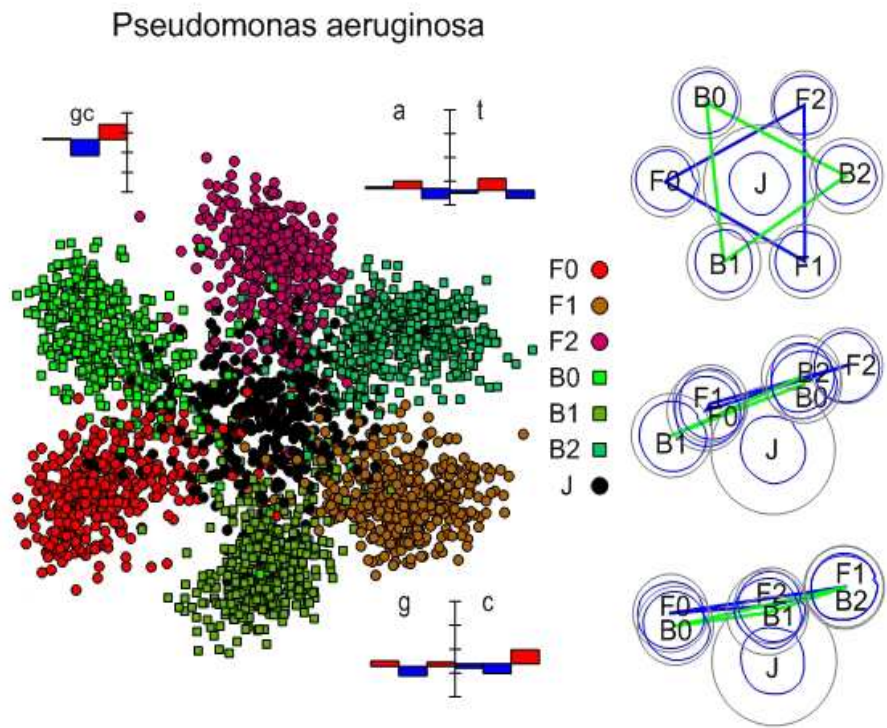}
}

\label{firstexample} \caption{Seven cluster structure of {\it
Pseudomonas aeruginosa} genomic sequence (G+C-content 67\%). On
the left pane the PCA plot of data distribution is shown. The
colors specify a frameshift, black circles correspond to
non-coding regions. On the right pane the structure is presented
in a schematic way, in three projections (first and second
principal components on the top, first and third in the middle,
second and third in the bottom), with ``radii'' of the clusters
schematically visualized. The diagrams show the codon
position-specific nucleotide frequencies (right top and right
bottom) as deviations from the average nucleotide frequency and
codon position-specific G+C-content (left top).}

\end{figure}

One example of the observed structure is shown on
Fig.\ref{firstexample}. In short, this is a PCA plot of the point
distribution. Referring for the details of the visualization to
the Methods section, we stop now on basic properties of the
structure. First, it consists of 7 clusters.This fact is rather
natural. Indeed, we clip fragments only from the forward strand
and every fragment can contain 1) piece of coding region from the
forward strand, with three possible shifts relatively to the first
fragment position; 2) coding information from the backward strand,
with three possible frameshifts; 3) non-coding region; 4) mix of
coding and non-coding information: these fragments introduce noise
in our distribution, but their relative concentration is not high.
Second, the structure is well pronounced, the clusters are
separated from each other with visible gaps. This means that most
of learning (and even self-learning) techniques aiming at
separation of the clusters from each other will work very well,
which is the case for bacterial gene-finders that have performance
more than 90\% in most cases (for recent overview, see
\cite{Mathe02}). Third, the structure is well represented by a
3D-plot (in this case it is even almost flat, i.e. 2D). Forth, it
is indeed has symmetric and appealing flower-like pattern, hinting
at there should be a symmetry in our statistics governing the
pattern formation.

In this paper we show how the structure depends on very general
properties of genomic sequence and show that it almost uniquely
depends on a single parameter: the genomic G+C content. Also,
based on analysis of 143 completely sequenced genomes, available
in Genbank in August 2004, we describe four ``pure'' types of the
structure observed in bacterial genomic sequences.

The outline of the paper is the following: first we introduce
phaseshift and complementary reverse operators, helping to
describe the structure, then we show that in nature we have almost
one-dimensional set of triplet distributions. After that we
explain the properties of the 7-cluster structure and describe
four ``pure'' types of the structures, observed for bacterial
genomes. The paper is finalized by the description of the methods
utilized and conclusion.

\section{Phaseshifts in triplet distributions}

Let us denote frequencies of non-overlapping triplets for a given
fragment as $f_{ijk}$, where $i,j,k\in\{A,C,G,T\}$, such as
$f_{ACT}$, for example, is a relative (normalized) frequency of
$ACT$ triplet.

One can introduce such natural operations over frequency
distribution as\textit{ phase shifts} P$^{{\rm (}{\rm 1}{\rm )}}$
, P$^{{\rm (}{\rm 2}{\rm )}}$ and \textit{complementary reversion}
C$^{{\rm R}}$:

\begin{equation}
\label{opdefinitions}
P^{(1)}f_{ijk} \equiv {\sum\limits_{l,m,n}
{f_{lij} f_{kmn}}}  , \quad P^{(2)}f_{ijk} \equiv
{\sum\limits_{l,m,n} {f_{lmi} f_{jkn}}}  , \quad \hat {f}_{ijk}^{}
\equiv C^{R}f_{ijk} \equiv f_{\hat {k}\,\hat {j}\,\hat {i}} ,
\end{equation}

\noindent where $\hat {i}$ is complementary to $i$, i.e., $ \hat
{A} = T, \hat {C}=~G, $ etc.

The phase-shift operator $P^{(n)}$ calculates a new triplet
distribution, counted with a frame-shift on $n$ positions, in the
hypothesis that no correlations exist in the order of triplets in
the initial phase. Complementary reversion constructs the
distribution of codons from a coding region in the complementary
strand, counted from the forward strand (``shadow'' frequency
distribution).

Phaseshift operators approximate the shifted triplet frequency as
superposition of a phase-specific nucleotide frequency and a
diplet frequency. This can be better understood if we rewrite
definitions (\ref{opdefinitions}) in the following way:

\begin{equation}
P^{(1)}f_{ijk} \equiv \sum_{l}f_{lij} \sum_{m,n} f_{kmn} \equiv
d^{(right)}_{ij}p^{(1)}_k, \quad P^{(2)} f_{ijk} \equiv \sum_{l,m}
f_{lmi} \sum_{n}f_{jkn} \equiv p^{(3)}_{i}d^{(left)}_{jk}
\end{equation}

We introduce the notion of \textit{mean-field} (or \textit{context
free}) approximation of the triplet distributions in the following
way:

\begin{equation}
\label{meanfield}
m_{ijk} = p_{i}^{(1)} p_{j}^{(2)} p_{k}^{(3)} ,
\quad p_{i}^{(1)} = {\sum\limits_{jk} {f_{ijk}}}  , \quad
p_{j}^{(2)} = {\sum\limits_{ik} {f_{ijk}}}  , \quad p_{k}^{(3)} =
{\sum\limits_{ij} {f_{ijk}}}  ,
\end{equation}

\noindent i.e. the mean-field approximation is the distribution
constructed from the initial triplet distribution neglecting all
possible correlations in the order of nucleotides. The
$p_{i}^{(k)} $ are the frequencies of the $i$th nucleotide ($i \in
$\{A,C,G,T\}) at the $k$th position of a codon ($k$ = 1..3). In
this way we model the 64 frequencies of the triplet distribution
using only 12 frequencies of the three position-specific
nucleotide distributions. This approximation is widely used in the
literature (see, for example, \cite{Bernaola}). All triplet
distributions that can be represented in the form
(\ref{meanfield}) belong to a 12-dimensional curved manifold {\bf
M}, parametrized by 12 frequencies $p_i^{(k)}$. The manifold is
embedded in the 64-dimensional space of all possible triplet
distributions {\bf T}\footnote{The normalization equality
$\sum_{ijk}f_{ijk} = 1$ makes all distributions to form a standard
63-dimensional simplex in $R^{64}$. For {\bf M} one has 3
independent normalizations: $\sum_{i}p^{(k)}_{i} = 1,\quad k =
1..3$, these equalities distinguish a 9-dimensional set (image of
the product of three 3-dimensional standard simplexes) in {\bf M},
where all normalized distributions are located.}.

It is easy to understand that any phase-shift for $m_{ijk}$ only
rotates the upper (position) indexes:

\begin{equation}
P^{(1)}m_{ijk} = p_{i}^{(2)} p_{j}^{(3)} p_{k}^{(1)} = m_{kij},
\quad P^{(2)}m_{ijk} = \left( {P^{(1)}} \right)^{2}m_{ijk} =
p_{i}^{(3)} p_{j}^{(1)} p_{k}^{(2)} = m_{jki}.
\end{equation}

Also it is worth noticing that applying the $P^{{\rm (}{\rm 1})}$
(or $P^{{\rm (}{\rm 2}{\rm )}})$ operator several times to the
initial triplet distribution we get the ($p_{i}^{(1)} p_{j}^{(2)}
p_{k}^{(3)} $,$p_{i}^{(2)} p_{j}^{(3)} p_{k}^{(1)} $,$p_{i}^{(3)}
p_{j}^{(1)} p_{k}^{(2)} )$ triangle:

\begin{equation}
\left( {P^{(1)}} \right)^{3}f_{ijk} = m_{ijk} .
\end{equation}

Operator $\left( {P^{(1)}} \right)^{3}$ acts as a projector
operator from full 64-dimensional distribution space $\bf{T}$ onto
the 12-dimensional manifold {\bf M}:

\begin{equation}
\left( {P^{(1)}} \right)^{3}: \bf{T} \rightarrow \bf{M}.
\end{equation}

On the manifold $\bf{M}$ of all possible $m_{ijk}$ we have
$P^{(2)}=(P^{(1)})^2$, therefore, there are only two operators:
phaseshift $P:Pm_{ijk} = m_{jki}$ and reversion $C:Cm_{ijk} =
m_{\hat{k}\hat{j}\hat{i}}$. There are following basic equalities:

\begin{equation}
P^3 = 1, \quad C^2 = 1, \quad PCP = C.
\end{equation}

Let us consider a point $m$ on $\bf{M}$. It corresponds to a  set
of 12 phase-specific nucleotide frequencies $p^{(1)}_i$, \quad
$p^{(2)}_i$ and $p^{(3)}_i$, $i\in\{A,C,G,T\}$. Applying operators
$P$ and $C$ in all possible combinations we obtain {\it an orbit}
on $\bf{M}$, consisting of 6 points: $m, Pm, P^2m, Cm, PCm,
P^2Cm$. Theoretically, some points can coincide, but only in such
a way that the resulting orbit will consist of 1 (fully
degenerated case), 3 (partially degenerated case) or 6
(non-degenerated case) points. The fully degenerated case
corresponds to the triplet distribution with the highest possible
entropy among all distributions with the same nucleotide
composition:

\begin{equation}
f_{ijk} = p_ip_jp_k, \quad p_i =
\sum_{mn}\frac{(f_{imn}+f_{mni}+f_{nim})}{3}.
\end{equation}

This distribution (completely random) is described by 4 nucleotide
frequencies, with any information about position in the triplet
lost. In our {\bf T} space it is a 3-dimensional (due to
normalization equality) simplex on {\bf M}. For bacterial genomes
this distribution can serve as an approximate (zero-order
accuracy) model for triplet composition in non-coding regions.

Now let us consider triplet distributions corresponding to the
codon usage of bacterial genomes, i.e. the subset of {\it
naturally occured triplet distributions}. It was found that they
could be reasonably well approximated by their mean-field
distributions, i.e. they are located close to $\bf{M}$. Moreover,
in the next section we show that in nature, for 143 completely
sequenced bacterial genomes, the $m_{ijk}$ distributions are
tightly located along one-dimensional curve on $\bf{M}$. The curve
can be parametrized by the genomic G+C-content.

\section{One-dimensional model of codon usage}

Twelve dependencies $p^{(1)}_i(GC)$, $p^{(2)}_i(GC)$ and
$p^{(3)}_i(GC)$, $i\in\{A,C,G,T\}$, where $GC$ is genomic
G+C-content, are presented on Fig.2(a-d) for 143 fully sequenced
bacterial genomes available in Genbank in August, 2004. These
dependencies are almost linear. This fact, despite it's
simplicity, was not explicitly demonstrated before. The numerous
results on the structure of codon usage described previously in
literature (see, for example,
\cite{Zhang91},\cite{Zhang94},\cite{Trifonov87}) are in agreement
with this picture.

Fig.2e demonstrates that the codon position-specific G+C-content
is a linear function of genomic G+C-content, in each position. In
\cite{Zhang03} analogous results were shown for 33 completed
genomes.

For our purposes, it is important to notice that for the genomes
with G+C-content higher than $\sim 60\%$ there is the same
well-defined structure in their codon usage: the G+C-content in
the first codon position is close to the average of three values,
the second is lower than the average and the third is essentially
higher than the average. This pattern can be denoted in the form
of simple {\it GC-signature}: $0-+$. In the next section we
develop more complicated signature to classify 7-cluster
structures, corresponding to the orbits, generated by $P$ and $C$
operators in a set of genomic fragments of 300-400bp length.

The linear functions, describing codon usage, are slightly
different for eubacteria and achaea genomes. Significant
deviations are observed for $p_A^{(1)}, p_C^{(1)}, p_G^{(3)},
p_G^{(1)}, p_T^{(3)}$ functions. For the others, the dependencies
are statistically indistinguishable.

One surprising conclusion follows from Fig.2: if we take the set
of triplet frequencies, occurred in nature and corresponding to
the codon usage of bacterial genomes, then in the 12-dimensional
space of codon position-specific nucleotide frequencies this set
appears almost as a straight line (more precisely, two close
lines, one for eubacteria and the other for archaea). If we look
at this picture from the 64-dimensional space of triplet
frequencies {\bf T}, then one sees that the distributions are
located close to the curved {\bf M} manifold of the mean-field
approximations, embedded in the space. As a result, when analyzing
the structure of the distribution of bacterial codon usage, one
detects that the points are located along two curves. These curves
coincide at their AT-rich ends and diverge at GC-rich ends. Moving
along these curves one meets all bacterial genomes. Genomes with
close G+C-content generally have close codon usage. Many evidences
for this structure were reported in studies on multivariate
analysis of bacterial codon usage (for example, see Figure 6 from
\cite{Lobry03}), but here we first explain the most general
structure of these representations in a clear and formal way.

\begin{figure}

\centering{
\begin{tabular}{cc}

 \includegraphics[width=70mm, height=70mm]{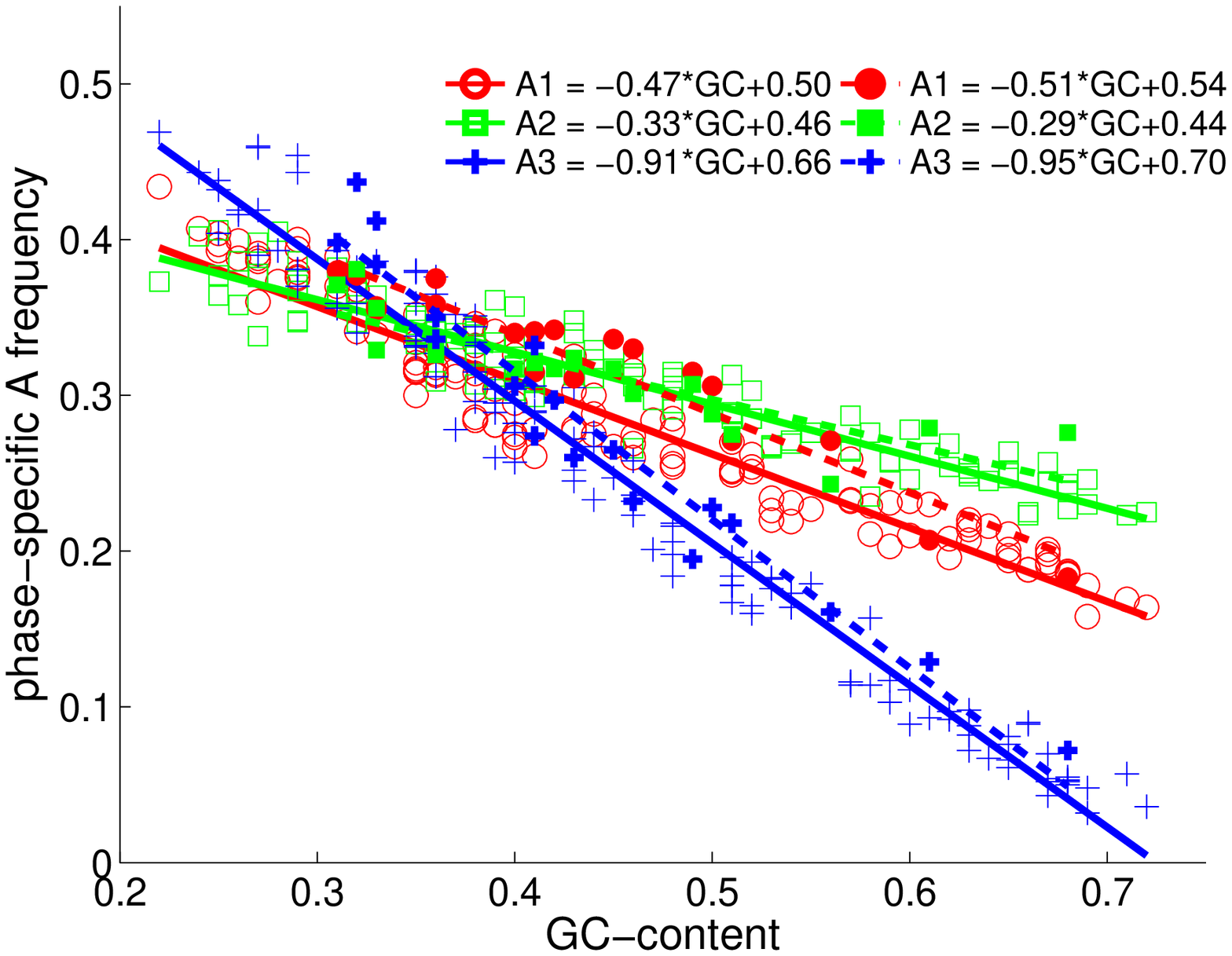}a)
 &
\includegraphics[width=70mm, height=70mm]{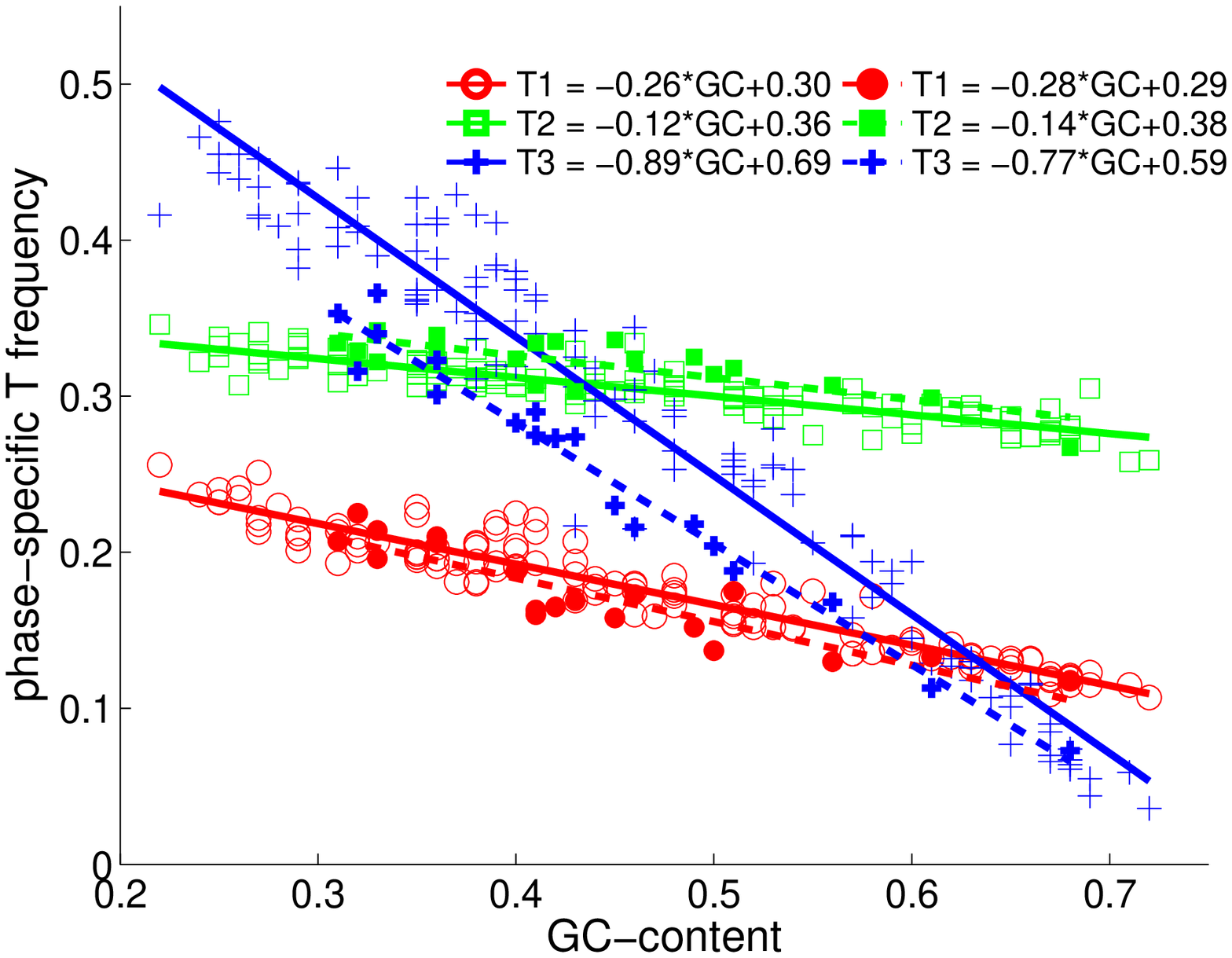}b)
  \\
\includegraphics[width=70mm, height=70mm]{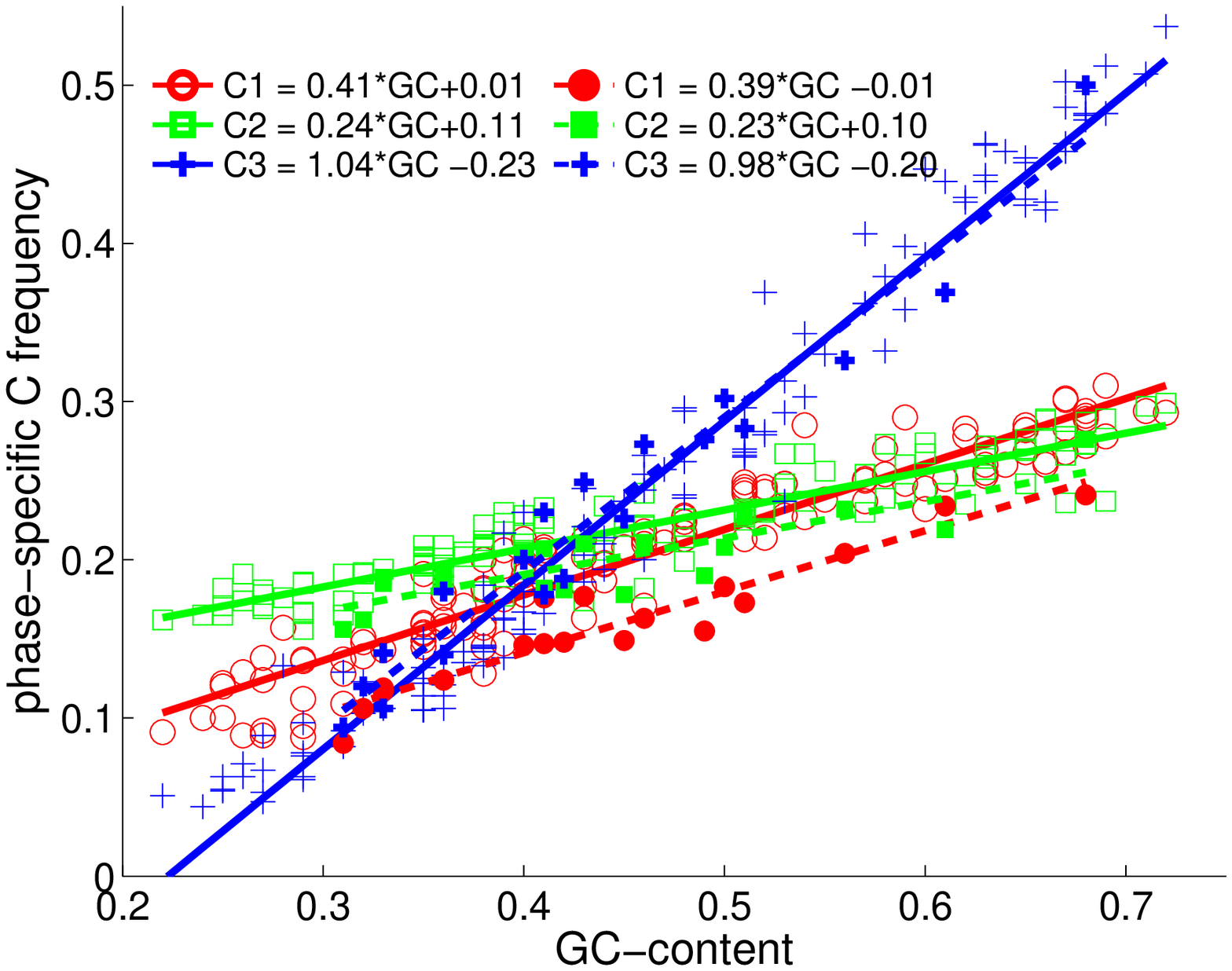}c)
 &
\includegraphics[width=70mm, height=70mm]{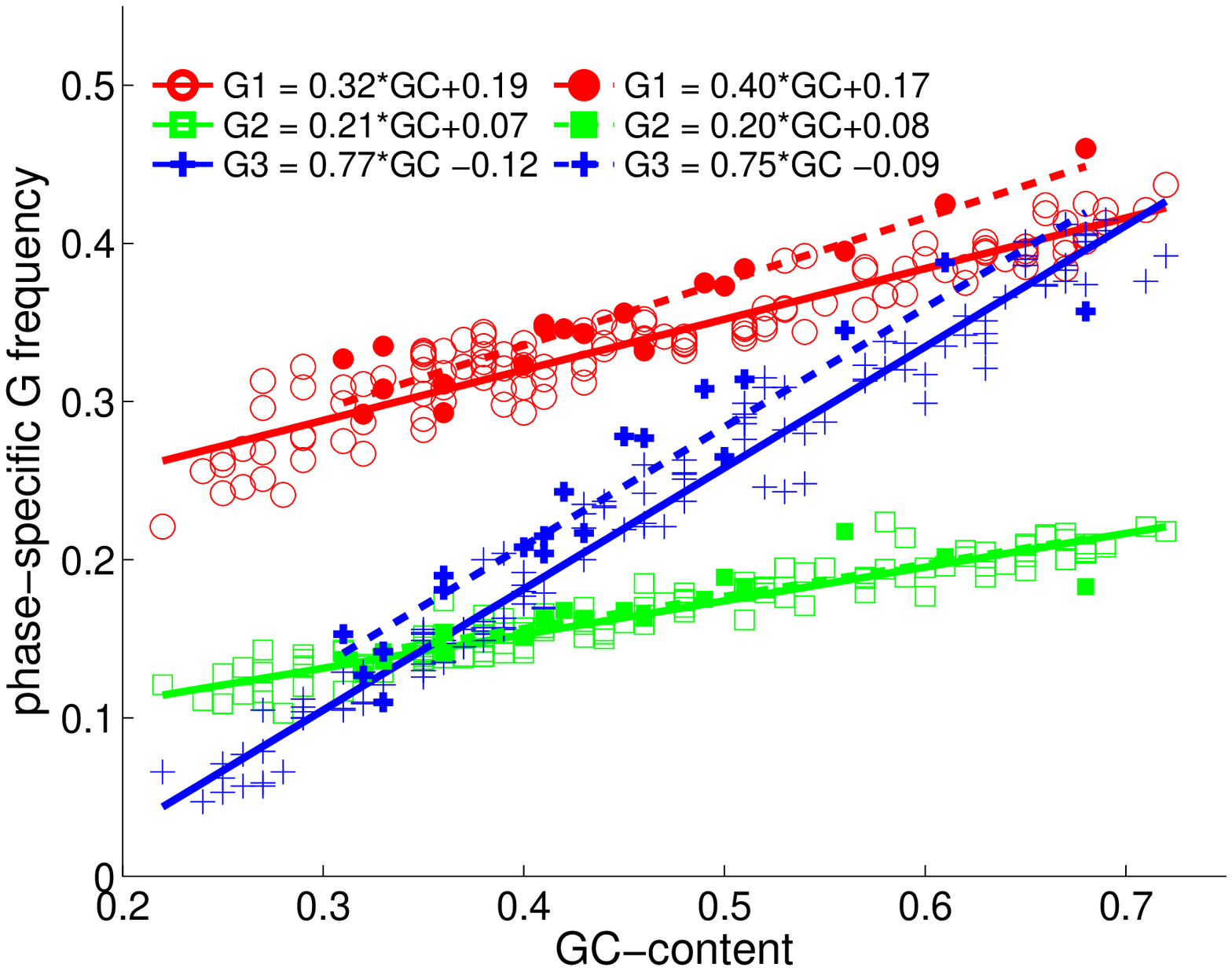}d)
 \\

\multicolumn{2}{c}{
\includegraphics[width=70mm, height=70mm]{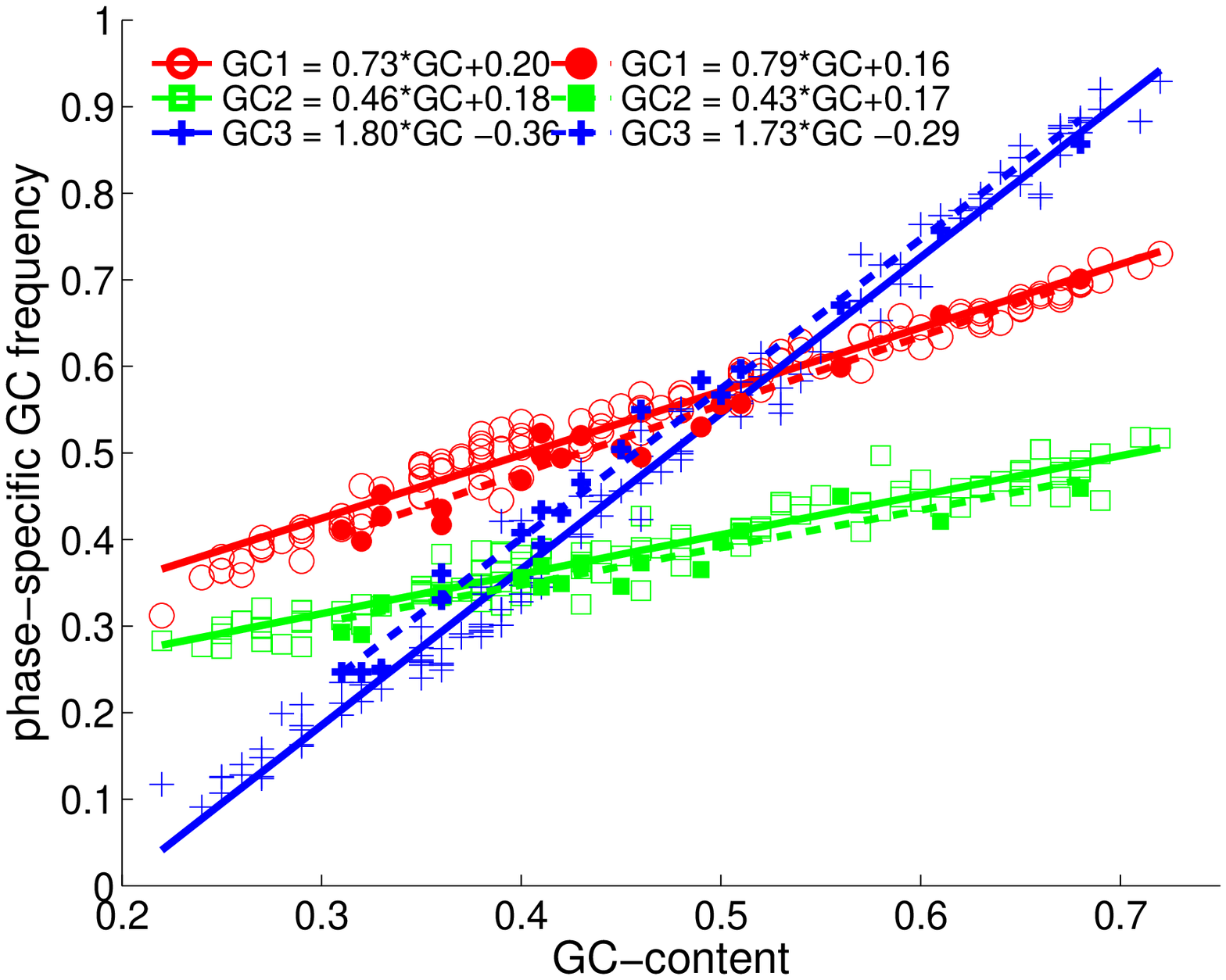}e)
}
\\
\end{tabular}
}\label{modelgraph} \caption{Codon position-specific nucleotide
frequencies (a-d) and codon position-specific GC-content (e).
Solid line and empty points correspond to 124 completed
eubacterial genomes, broken line and filled points correspond to
19 completed archaeal genomes. }
\end{figure}

\section{Properties and types of the 7-cluster structure}

In the paper \cite{Zhang03} the authors claim that the codon
position-specific nucleotide frequencies (represented as
Z-coordinates) in GC-rich genomes show flower-like cluster
structure, and the phenomenon is not observed in other genomes.
Here we explain the phenomenon and demonstrate other types of
structures observed in genomes and that the type of the structure
is related to the pattern of symmetric properties of codon usage.

First of all, we point out to the fact that the space used in
\cite{Zhang03} is a specific projection from 64-dimensional space
of triplet frequencies. The 9-dimensional phenomenon is also can
be observed in 64-dimensional space and in 12-dimensional space of
codon position-specific nucleotide frequencies.

Let us consider the context free approximation of codon usage
introduced above:

\begin{equation}
m_{ijk} = p^{(1)}_{i}p^{(2)}_{j}p^{(3)}_{k}
\end{equation}

and consider 3D space with the following coordinates:

\begin{equation}
x = p^{(1)}_G+p^{(1)}_C - f_{GC}, y = p^{(2)}_G+p^{(2)}_C -
f_{GC}, z = p^{(3)}_G+p^{(3)}_C - f_{GC}
\end{equation}

In fact, x, y and z are deviations of GC-content in the first,
second and the third position from average GC-content $f_{GC}$ of
coding regions. In all GC-rich genomes (starting from
$f_{GC}>60\%$) their codon usage context-free approximation has
the following structure (see Fig.2e): $x\approx 0, y<0, z>0$. We
can denote this pattern as $0-+$. Applying phaseshift and reverse
operators defined above (notice that $C$ operator does not change
G+C-content, it only reverses the signture), we obtain the
following orbit: $\{0-+,-+0,+0-\}$ and $\{+-0,-0+,0+-\}$. If now
we consider a 3D grid consisting of 27 nodes as shown on Fig.3,
corresponding to all possible patterns (GC-signatures), then it is
easy to understand that the orbit corresponds to the points of
where the grid is cross-sectioned by a plane, coming through the
$000$ point perpendicular to the $\{---,+++\}$ diagonal. It is
well known fact that in this situation the form of the
intersection is a regular hexagon. The $000$ point in our picture
corresponds to the center of the non-coding cluster (this is the
fully degenerated distribution described above), where all phases
have been mixed. The $\{---,+++\}$ diagonal corresponds to the
direction of the fastest G+C-content increase. Hence, this model
explains the following features of the flower-like structure
observed in GC-rich ($G+C>60\%$) genomes:

1) In the 64-dimensional space the centers of clusters are
situated close to a distinguished 2D-plane, forming regular
hexagonal structure.

2) The third principal component (perpendicular to the cluster
plane) is the direction of G+C-content increase (i.e., the
gradient of G+C-content linear function, defined in the
64-dimensional triplet space).

In most flower-like structures the cluster that corresponds to the
non-coding regions is slightly displaced in the direction
perpendicular to the main cluster plane. This happens because
G+C-content of non-coding regions is generally slightly lower than
of coding regions.

\begin{figure}
\label{cube} \centering{
\includegraphics[width=80mm, height=80mm]{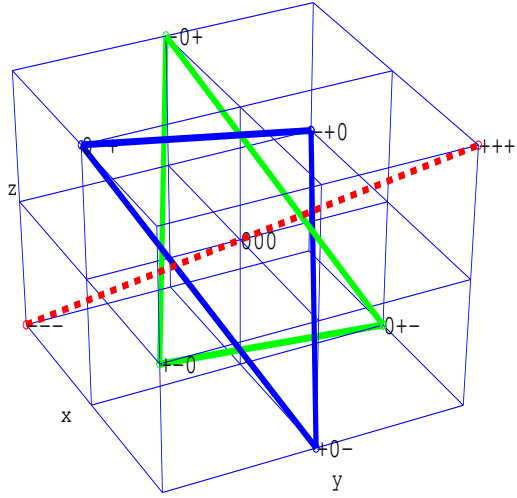}
\caption{Model of the flower-like cluster structure. Broken line
corresponds to the direction of the fastest G+C-content increase.}
}
\end{figure}

Now let us consider general case of genome with any given genomic
G+C-content. The type of the 7-cluster structure depends on values
of 12 functions $p^{(1)}_i, p^{(2)}_i, p^{(3)}_i$, $i\in
\{A,C,G,T\}$. Applying phasesift and reverse operators, one
obtains an orbit which serves as a skeleton of the cluster
structure. The orbit structure reflects symmetries in the set of
values of these 12 functions with respect to the $P$ and $C$
operators.

We describe these symmetries in the following simplified manner.
Let us order the 12 values in the form of $6\times 2$ table:

\begin{equation}
s_{ij} =
\begin{array}{c}
  p^{(1)}_A p^{(2)}_A p^{(3)}_A p^{(1)}_T p^{(2)}_T
p^{(3)}_T \\
  p^{(1)}_G p^{(2)}_G p^{(3)}_G p^{(1)}_C p^{(2)}_C
p^{(3)}_C
\end{array}
\end{equation}

Then the reverse operator $C$ simply reads the table from right to
the left:

\begin{equation}
Cs_{ij} =
\begin{array}{c}
  p^{(3)}_T p^{(2)}_T p^{(1)}_T p^{(3)}_A p^{(2)}_A
p^{(1)}_A \\
  p^{(3)}_C p^{(2)}_C p^{(1)}_C p^{(3)}_G p^{(2)}_G
p^{(1)}_G
\end{array}.
\end{equation}

The phaseshift operator $P$ rotates the values in the table by
threes, for every letter:

\begin{equation}
Ps_{ij} =
\begin{array}{c}
  p^{(3)}_A p^{(1)}_A p^{(2)}_A p^{(3)}_T p^{(1)}_T
p^{(2)}_T \\
  p^{(3)}_G p^{(1)}_G p^{(2)}_G p^{(3)}_C p^{(1)}_C
p^{(2)}_C
\end{array}.
\end{equation}

We reduce the description of $s$ in the following way: every entry
in the table is substituted by ``$+$'', ``$-$'' and ``$0$'', if
the corresponding value is bigger then the average over the same
letter frequencies, smaller or in the
$[average-0.01;average+0.01]$ interval respectively. For example,
for a set of frequencies $p^{(1)}_A=0.3$,$p^{(2)}_A=0.5$,
$p^{(3)}_A=0.401$, we substitute $p^{(1)}_A\rightarrow -$,
$p^{(2)}_A\rightarrow +$, $p^{(3)}_A\rightarrow 0$. We call
``signature'' the new table $\hat{s}_{ij}$ with reduced
description.

Using linear folmulas from the Fig.2(a-d) and calculating the
$\hat{s}_{ij}$ tables for the range $[0.2;0.8]$ of G+C-content, we
obtain 19 possible signatures in the intervals of genomic
G+C-content, listed in Table \ref{allsigns}.

\begin{table}
  \centering{

\begin{tabular}{|l|l|l|}

  $\begin{array}{c}  \texttt{--+--+}\\\texttt{+--++-}   \end{array} [0.200;0.255) $
  & $\begin{array}{c} \texttt{000-++}\\\texttt{+--0+-}   \end{array} [0.331;0.373)$ &
  $\begin{array}{c} \texttt{0+--++}\\\texttt{+---0+}   \end{array} [0.434;0.482)$  \\
    &&\\
  $\begin{array}{c} \texttt{--+--+}\\\texttt{+--0+-}   \end{array} [0.255;0.265) $
  & $\begin{array}{c} \texttt{0+0-++}\\\texttt{+--0+-}   \end{array} [0.373;0.385)$ &
$\begin{array}{c} \texttt{0+--++}\\\texttt{+-0-0+}   \end{array}
[0.482;0.487)$ \\
   &&\\
  $\begin{array}{c} \texttt{--+-0+}\\\texttt{+--0+-}   \end{array} [0.265;0.289) $
  & $\begin{array}{c} \texttt{0+--++}\\\texttt{+--0+-}   \end{array} [0.385;0.388)$ &
  $\begin{array}{c} \texttt{0+--++}\\\texttt{+-0--+}   \end{array} [0.487;0.502)$\\
   &&\\
  $\begin{array}{c} \texttt{-0+-0+}\\\texttt{+--0+-}   \end{array} [0.289;0.316) $
  & $\begin{array}{c} \texttt{0+--++}\\\texttt{+---+-}   \end{array} [0.388;0.391)$ &
    $\begin{array}{c} \texttt{0+--+0}\\\texttt{+-0--+}   \end{array} [0.502;0.515)$\\
   &&\\
  $\begin{array}{c} \texttt{00+-0+}\\\texttt{+--0+-}   \end{array} [0.316;0.326) $
  & $\begin{array}{c} \texttt{0+--++}\\\texttt{+---+0}   \end{array} [0.391;0.424)$ &
    $\begin{array}{c} \texttt{++--+0}\\\texttt{+-0--+}   \end{array} [0.515;0.542)$\\
   &&\\
  $\begin{array}{c} \texttt{000-0+}\\\texttt{+--0+-}   \end{array} [0.326;0.331) $
  & $\begin{array}{c} \texttt{0+--++}\\\texttt{+---00}   \end{array} [0.424;0.434)$ &
    $\begin{array}{c} \texttt{++--+0}\\\texttt{+-+--+}   \end{array} [0.542;0.545)$  \\
   &&\\
  && $\begin{array}{c} \texttt{++--+-}\\\texttt{+-+--+}   \end{array} [0.545;0.800)$
    \\

\end{tabular}

  }
  \caption{Nineteen possible signatures for one-dimensional codon
usage model}\label{allsigns}
\end{table}

There are 67 different signatures observed for really occurred
$p_i^{(k)}$-values for 143 genomes considered in this work (see
our web-site \cite{web} with supplementary materials). Most of
them differ from the signature in Table \ref{allsigns} with
corresponding G+C value only by changing one of the ``+'' or
``$-$'' for ``0'' or vise versa.

From Table \ref{allsigns} one can see that the only conserved
positions, independent on genomic G+C-content for the interval
$[0.20;0.80]$ are $p^{(1)}_T$ (always ``$-$''), $p^{(1)}_G$
(always ``+''), $p^{(2)}_G$ (always ``$-$''). This holds true also
for all really observed signatures. This observation confirms
already known ``invariants'' of codon usage described in
\cite{Zhang91},\cite{Zhang94},\cite{Trifonov87}.

Let us look at several typical examples. All genomes with genomic
G+C-content higher then $60\%$ have the following genomic
signature:

\begin{equation}
\hat{s}_{ij}(GC>60\%) =
\begin{array}{c}
\texttt{++--+-}\\\texttt{+-+--+}
\end{array}.
\end{equation}

This signature uniformly reflects the previously mentioned
$GC$-signature (``$0-+$''): pairs $p_A^{(1)}$,$p_T^{(1)}$ and
$p_G^{(1)}$,$p_C^{(1)}$ compensate the signs of each other to give
``0'' in the first position of $GC$-signature, while in the second
position we have ``+'' for A and T and ``$-$'' for G and C, and
vice versa for the third position. As a result, we obtain the
flower-like structure. On Fig.4 we visualize the orbit for {\it
Streptomyces coelicolor}, genome with high G+C-content: 72\%.
Together with the orbit we visualize the distance matrix for the
skeleton, where the distances are calculated in the full
64-dimensional triplet frequency space ${\bf T}$. Black color on
the plot corresponds to zero distance, white for the biggest value
in the matrix. The most informative $3\times 3$ block of the
matrix is in the left bottom corner (or top right, by symmetry):
it describes mutual distances between the vertexes of two skeleton
triangles. The left top and right bottom $3\times 3$ blocks
contain equal values, since the sides of the triangles have the
same length.

\begin{figure}
\label{examples} \centering{
\begin{tabular}{cc}

 \includegraphics[width=70mm, height=70mm]{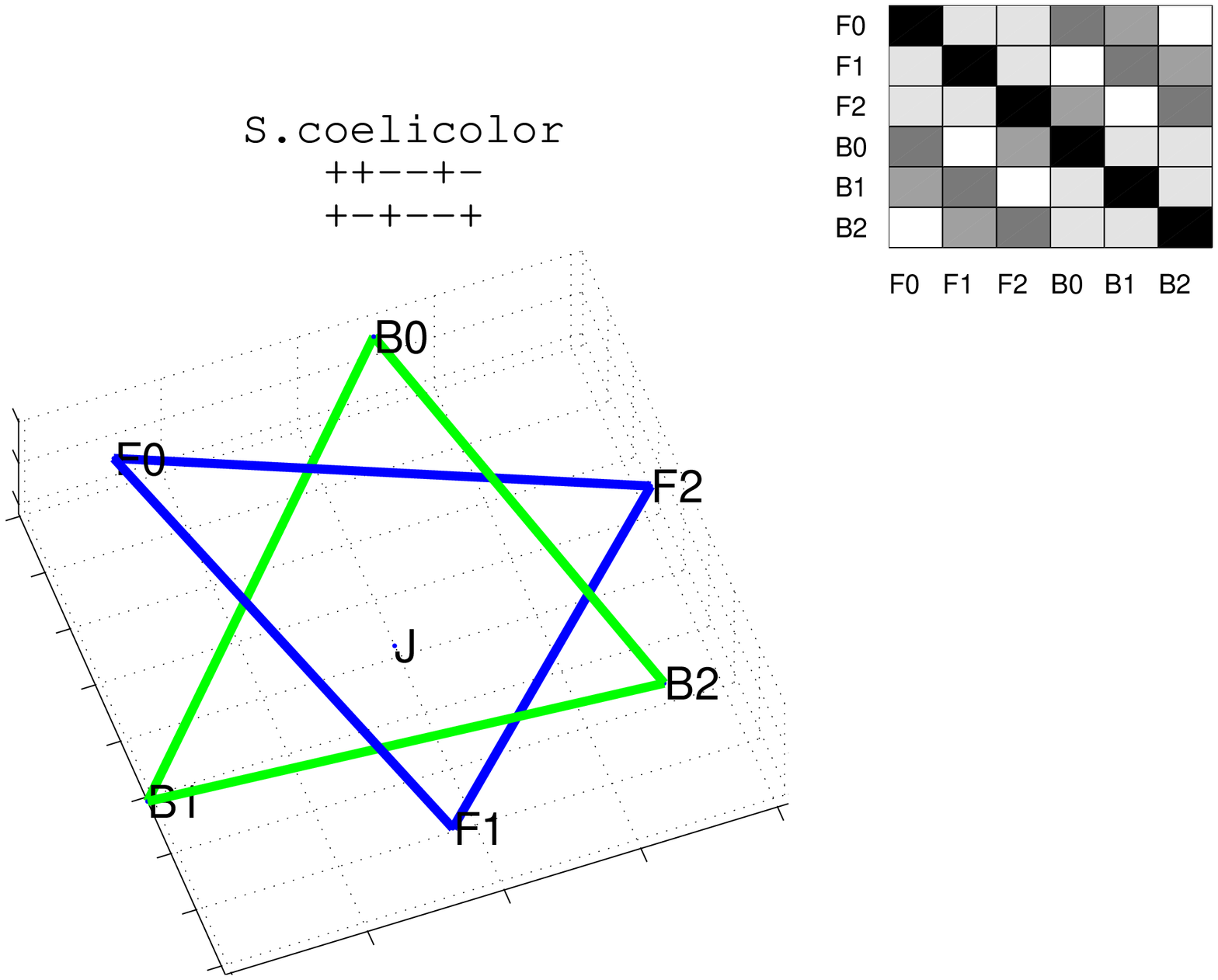}a)
 &
\includegraphics[width=70mm, height=70mm]{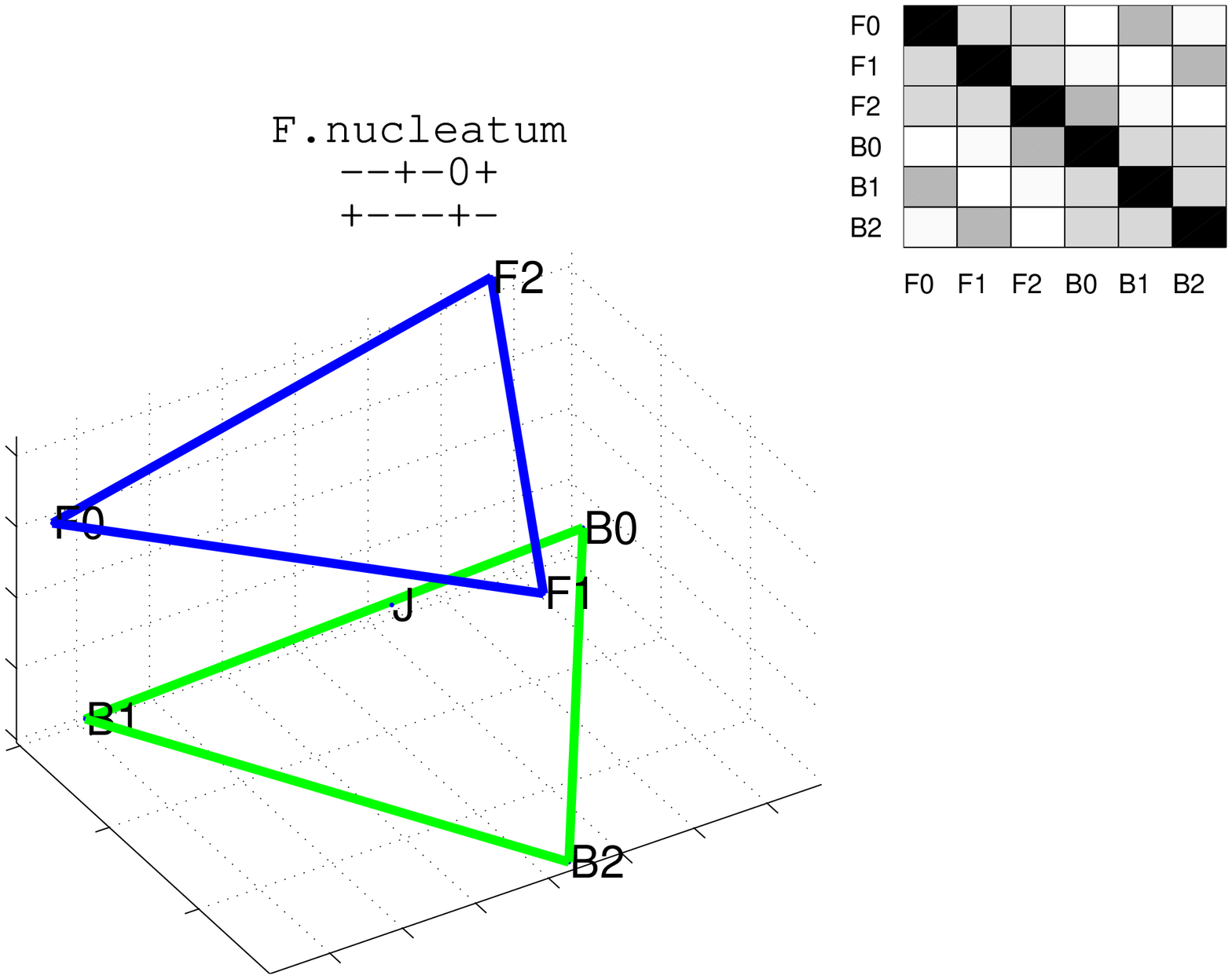}b)
  \\
\includegraphics[width=70mm, height=70mm]{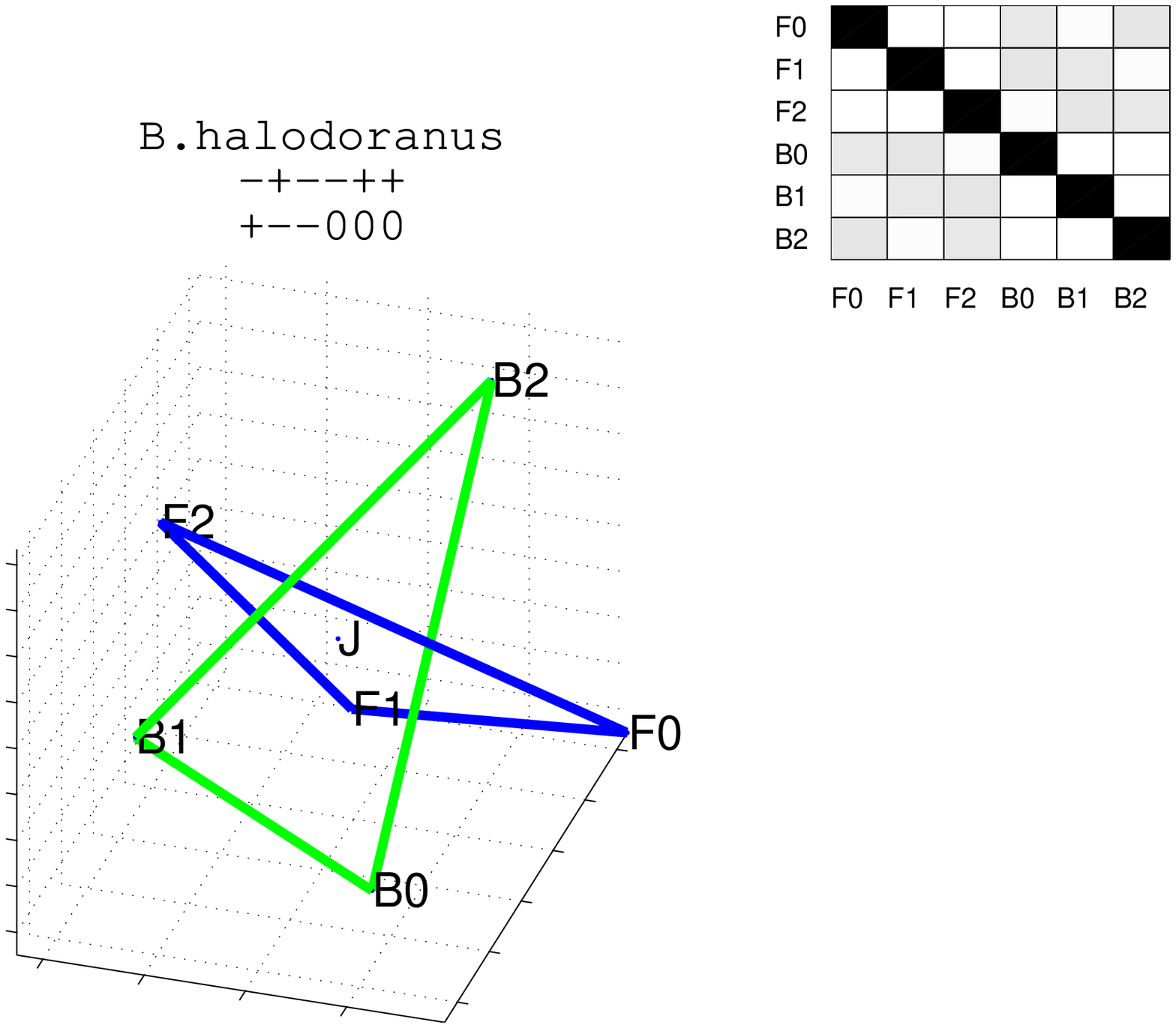}c)
 &
\includegraphics[width=70mm, height=70mm]{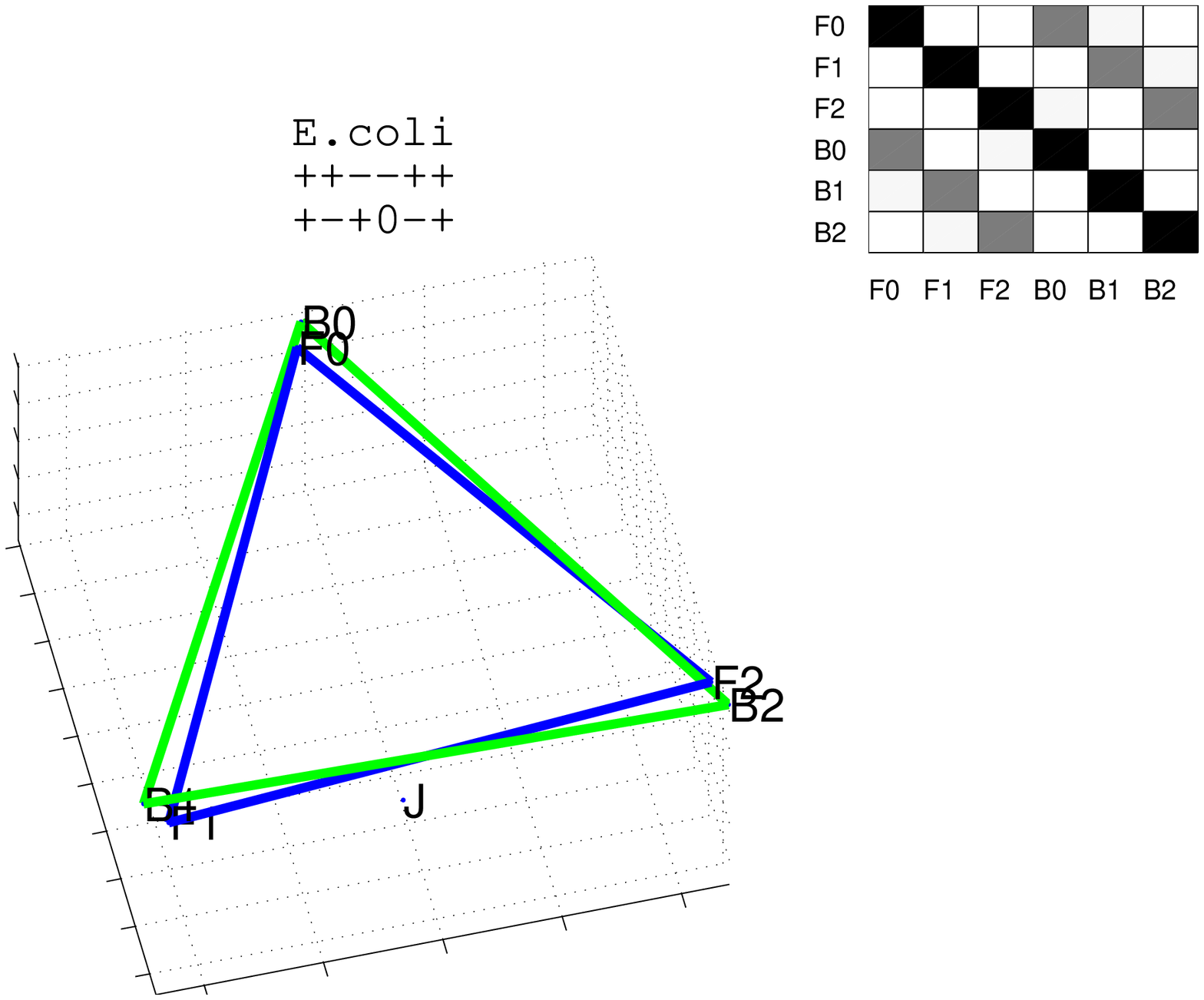}d)

\end{tabular}
}

\caption{Four typical examples of the 7-cluster structure: a)
genome of S.coelicolor (GC=72\%), flower-like structure; b) genome
of F.nucleatum(GC=27\%), ``parallel triangles''; c) B.halodurans
(GC=44\%), ``perpendicular triangles''; d) E.coli (GC=51\%),
degenerated case. }
\end{figure}

Our second example is genome of {\it Fusobacterium nucleatum}
(AT-rich genome, G+C-content is 27\%), Fig.4b. The signature is

\begin{equation}
\hat{s}_{ij}(F.nucleatum) =
\begin{array}{c}
\texttt{--+-0+}\\\texttt{+---+-}
\end{array}.
\end{equation}

This pattern, commonly observed in AT-rich genomes, can be called
``parallel triangles''. Notice that two parallel triangles are
rotated with respect to their corresponding phase labels: the F0
vertex is located in front of the B1 vertex.

The third example is genome of {\it Bacillus halodurans}
(G+C-content is 44\%):

\begin{equation}
\hat{s}_{ij}(B.halodurans) =
\begin{array}{c}
\texttt{-+--++}\\\texttt{+--000}
\end{array}.
\end{equation}

We refer to this pattern as ``perpendicular triangles''. Another
example of the pattern is genome of {\it Bacillus subtilis}. All
non-diagonal distances in the distance matrix have in this case
approximately the same value. This structure can be easily
recognised from it's signature: the second row has three zeros
while the first one is almost {\it palindromic}. As we will see in
the next example, palindromic rows in the signature (or such that
can be made palindromic applying the phaseshift $P$ operator) make
zero contribution to the diagonal of the ``inter-triangle'' part
of the distance matrix. This is easy to understand, because the
reverse operator $C$ reads the signature from right to the left.
The rows with three zeros in different phase positions (when, for
example, the phase specific nucleotide frequencies for one letter
are equal to their average, as happened in this case) give
approximately equal contribution to every value in the
``inter-triangle'' part of the distance matrix. The resulting
matrix corresponds to the ``perpendicular triangles'' pattern. We
should notice that the distance matrix showed on Fig.4c can not be
effectively represented as a distribution of 6 points in 3D. Thus
the ``perpendicular triangles'' structure shown on Fig.4c is only
an approximate picture, the real configuration is almost
6-dimensional, due to the distance matrix symmetry.

In the region of G+C-content about 51\% we observe a group of
genomes with almost palindromic signatures. One typical example is
the genome of {\it Escherichia coli}:

\begin{equation}
\hat{s}_{ij}(E.coli) =
\begin{array}{c}
\texttt{++--++}\\\texttt{+-+0-+}
\end{array}.
\end{equation}

The resulting pattern is a degenerated case: two triangles are
co-located, without phase label rotation (F0 is approximately in
the same point as B0). The distance matrix consists of 4 almost
identical $3\times 3$ blocks. As a result, we have situation, when
7-cluster structure effectively consists of only 4 clusters, one
for every pair F0-B0, F1-B1, F2-B2 and a non-coding cluster.

The same degenerated case but with rotation of labels
(F0-B1,F1-B2,F2-B0) is observed for some AT-rich genomes. For
example, for the genome of {\it Wigglesworthia brevipalpis}
(G+C-content equals 22\%) the signature

\begin{equation}
\hat{s}_{ij}(W.brevipalpis) =
\begin{array}{c}
\texttt{0-+-0+}\\\texttt{+---+-}
\end{array}
\end{equation}

becomes a perfect palindrom after applying the phaseshift
operator:

\begin{equation}
P\hat{s}_{ij}(W.brevipalpis) =
\begin{array}{c}
\texttt{-+00+-}\\\texttt{--++--}
\end{array}.
\end{equation}

One possible biological consequence (and even explanation) of this
degeneracy is existence of overlapping genes: in this case the
same codons can be used to encode proteins simultaneously in the
forward and backward strands on a regular basis (without
frameshift for G+C-content around 50\% and with a frameshift for
AT-rich genomes), with the same codon usage.

The four patterns are typical for triplet distributions of
bacterial genomes observed in nature. The other ones combine
features from these four ``pure'' types. In general, going along
the G+C-content scale, we meet first ``parallel triangles'' which
will transform gradually to ``perpendicular triangles''. On this
way one can even meet structures resembling flower-like type in
one of the 2D-projections, like for the genome of {\it
Helicobacter pylori} (see our web-site \cite{web} and in
\cite{Zinovyev02} for the illustration). Then the pattern goes to
the degenerated case with genomic G+C-content around 50\% and
signatures close to palindromic. After the degeneracy disappears,
the pairs F0-B0, F1-B1, F2-B2 diverge in the same 2D-plane and
after 55\% threshold in G+C-content we almost exclusively have the
flower-like structures. It is possible to browse the animated
scatters of 7-cluster structures observed for every of 143 genomes
on our web-site \cite{web}.

\section{Web-site on cluster structures in genomic word frequency distributions}

To make the images and graphs of 143 genomes 7-cluster structures
available for wide public, we established a web-site \cite{web}
for cluster structures in genomic word frequency distributions.

For the moment our database contains 143 completely sequenced
bacterial genomes and two types of cluster structures: the
7-cluster structure and the gene codon usage cluster structure.
When browsing the database, a user can look at animated
3D-representations of these multidimensional cluster structures.
For the description of the structures and the methods we refer the
reader to the ``intro'' and ``methods'' pages of the web-site.

Another possibility which is provided on our web-site is browsing
large-scale ``maps'' of various spaces where all 143 genomes can
be embedded simultaneously. One example is the codon usage map:
one point on the map is a genome, and close points correspond to
the genomes with close codon usage. In fact, this is the same
64-dimensional triplet frequency space, used for construction of
the 7-cluster structure, ``observed'' from a big distance. This
gives the following hierarchy of maps: general map of codon usage
in 143 genomes, then the 7-cluster structure of in-phase triplet
distributions, then the ``thin structure'' of every coding
cluster: gene codon usage map. Clicking on a genome at the first
map, the user ``zooms'' to it's more detailed representations.

We strongly believe that the information in the database will help
to advance existing tools for bacterial genomes analysis. Also it
can serve as rich illustrative material for those who study
sequence bioinformatics.

\section{Methods}

For visualization of the 7-cluster structure a data set for every
143 genome\footnote{We took only different species from GenBank,
choosing the first alphabetically available strain in the list.}
was prepared in the following way:

1) A Genbank file with completed genome was downloaded from the
Genbank FTP-site. Using BioJava package \cite{BioJava} the
complete sequence and the annotation was parsed. In the case when
the genome had two chromosomes, the sequences of both were
concantenated. The short sequences of plasmids were ignored.

2) Let $N$ be length of a given sequence $S$ and $S_i$ be a letter
in the $i$th position of $S$. We define a step size $p$ and a
fragment size $W$. For given $p$, $W$ and $k=1..[\frac{N-W}{p}]$
we clip a fragment of length $W$ in the sequence, centered in
$i=w/2+pk$. In every fragment we count frequencies of
non-overlapping triplets:

\begin{equation}
count(c_{i_1}c_{i_2}c_{i_3}) =
\sum_{i=0}^{W/3-1}comp(c_{i_1}c_{i_2}c_{i_3},S_{i*3+1}S_{i*3+2}S_{i*3+3}),
\end{equation}

where $comp(word1,word2)$ is a string comparison function that has
value 1 if $word1$ equals $word2$ and 0 otherwise. Here $c_i$ is a
letter from genetic alphabet ($c_1=A$, $c_2=C$, $c_3=G$, $c_4=T$).

For every fragment a frequency vector is defined:

\begin{equation}
X_i^{c_{i_1}c_{i_2}c_{i_3}} =
\frac{count(c_{i_1}c_{i_2}c_{i_3})}{\sum_{j_1,j_2j_3}{c_{j_1}c_{j_2}c_{j_3}}}
\end{equation}

All words, which contain non-standard letters like N, S, W, are
ignored.

The data set ${X_i^j}$, $j=1..[\frac{N-W}{p}]$ is normalized to
have unity standard deviation and zero mean.

3) We assign to the fragment a label accordingly to the Genbank
annotation of the CDS features (including hypothetical ones) that
include the center of the fragment. If the center is inside a CDS
feature then the reading frame and the strand of the CDS feature
are determined and the fragment is assigned one of
{F0,F1,F2,B0,B1,B2} label. In the second case the label is J.

4) Standard PCA-analysis is performed and the first three
principal components are calculated. They form form a
3D-orthonormal basis in the 64-dimensional space. Every point is
projected into the basis, thus we assign three coordinates for
every point .

To create a schema of 7-cluster structure (see
Fig.\ref{firstexample}) the following method was utilized. We
calculated the mean point $y^L$ for every subset with a given
label $L$.

For the set of centroids $y^{F0}$, $y^{F1}$, $y^{F2}$, $y^{B0}$,
$y^{B1}$, $y^{B2}$ a distance matrix of euclidean distances was
calculated and visualized using classical MDS.

To visualize the ``radii'' of the subsets, a mean squared distance
$d^{(p)}$ to the centroid $p$ was calculated (intraclass
dispersion). To visualize the value on 2D plane, we have to
introduce dimension correction factor, so the radius drawn on the
picture equals

\begin{equation}
r^{(p)}=\sqrt{\frac{2}{4^k}d^{(p)}}
\end{equation}

The form of the cluster is not always spherical and often
intersection of radii do not reflect real overlapping of classes
in high-dimensional space. To show how good the classes are
separated actually, we developed the following method for cluster
contour visualization. To create a contour for class $p$, we
calculate averages of all positive and negative projections on the
vectors connecting centroid $p$ and 6 other centroids $i=1..6$.

\begin{equation}
\bf{n}_i^p = \frac{y^i-y^p}{||y^i-y^p||},
\it{n}_i^p(X_k)=(X_k-y^p,\bf{n}^p_i)
\end{equation}

\begin{equation}
\it{f}^p_i=\frac{\sum_{\it{n}_i^p(X_k)>0}{\it{n}_i^p}}{\sum_{\it{n}_i^p(X_k)>0}{1}},
\it{b}^p_i=\frac{\sum_{\it{n}_i^p(X_k)<0}{\it{n}_i^p}}{\sum_{\it{n}_i^p(X_k)<0}{1}}
\end{equation}

Then, using the 2D MDS plot where every vector $(y^p)'$ has 2
coordinates, we put 12 points $t^f$, $t^b$ analogously.

\begin{equation}
(\bf{n}_i^p)' = \frac{(y^i)'-(y^p)'}{||(y^i)'-(y^p)'||},
\end{equation}

\begin{equation}
t^f_i = (y^i)'+f_i^p(n_i^p)', t^b_i = (y^i)'+b_i^p(n_i^p)', i=1..6
\end{equation}

Using a smoothing procedure in polar coordinates we create a
smooth contour approximating these 12 points.

\section{Discussion}

In this paper we prove the universal 7-cluster structure in
triplet distributions of bacterial genomes. Some hints at this
structure appeared long time ago, but only recently it was
explicitly demonstrated and studied.

The 7-cluster structure is the main source of sequence
heterogeneity (non-randomness) in the genomes of bacterial
genomes. In this sense, our 7 clusters is the basic model of
bacterial genome sequence. We demonstrated that there are four
basic ``pure'' types of this model, observed in nature: ``parallel
triangles'', ``perpendicular triangles'', degenerated case and the
flower-like type (see Fig.\ref{examples}).

To explain the properties and types of the structure, which occur
in natural bacterial genomic sequences, we studied 143 bacterial
genomes available in Genbank in August, 2004. We showed that,
surprisingly, the codon usage of the genomes can be very closely
approximated by a multi-linear function of their genomic
G+C-content (more precisely, by two similar functions, one for
eubacterial genomes and the other one for archaea). In the
64-dimensional space of all possible triplet distributions the
bacterial codon distributions are close to two curves, coinciding
at their AT-rich ends and diverging at their GC-rich ends. When
moving along these curves we meet all naturally occurred 7-cluster
structures in the following order: ``parallel triangles'' for the
AT-rich genomes (G+C-content is around 25\%), then ``perpendicular
triangles'' for G+C-content is around 35\%, switching gradually to
the degenerated case in the regions of GC=50\% and, finally, the
degeneracy is resolved in one plane leading to the flower-like
symmetric pattern (starting from GC=60\%). All these events can be
illustrated using the material from the web-site we established
\cite{web}.

The properties of the 7-cluster structure have natural
interpretations in the language of Hidden Markov Models. Locations
of clusters in multdimensional space correspond to in-state
transition probabilities, the way how clusters touch each other
reflects inter-state transition probabilities. Our clustering
approach is independent on the Hidden Markov Modeling, though can
serve as a source of information to adjust the learning
parameters.

In our paper we analyzed only triplet distributions. It is easy to
generalize our approach for longer (or shorter) words. In-phase
hexamers, for example, are characterized by the same 7-cluster
structure. However, our experience shows that the most of
information is contained in triplets: the correlations in the
order of codons are small and the formulas (\ref{opdefinitions})
work reasonably well. Other papers confirm this observation (see,
for example, \cite{Claverie98}, \cite{Gorban03}).

The subject of the paper has a lot of possible continuations.
There are several basic questions: how one can explain the
one-dimensional model of codon usage or why the signatures in the
middle of G+C-content scale have palindromic structures? There are
questions about how our model is connected with codon bias in
translationally biased genomes: the corresponding cluster
structure is the second hierarchical level or the ``thin
structure'' in every cluster of the 7-cluster structure (see, for
example, \cite{Carbone03}). Also the following question is
important: is it possible to detect and use the universal
7-cluster structure for higher eukaryotic genomes, where this
structure also exists (see \cite{Zinovyev02}), but is hidden by
the huge non-coding cluster?

The information about the 7-cluster structure can be readily
introduced into existing or new software for gene-prediction,
sequence alignment and genome classification.

\end{document}